\documentclass[12pt]{article}
\usepackage{amsfonts}
\usepackage{amsmath,amssymb}
\usepackage{stmaryrd}
\usepackage[dvips]{color}
\usepackage{fancybox}
\usepackage{bm}
\usepackage[dvips]{graphicx}
\usepackage{wrapfig}

\baselineskip=\normalbaselineskip

\begin{document} 
\titlepage
\setcounter{page}{0}

\numberwithin{equation}{section}
\renewcommand{\theequation}{\thesection.\arabic{equation}}
\def\natural{\mathbb{N}}
\newcommand{\Bra}[1]{\left\langle\, #1\,\right|}
\newcommand{\Ket}[1]{\left|\, #1\,\right\rangle}
\newcommand{\Bracket}[2]{\left\langle\, #1\,|\, #2\,\right\rangle}
\newcommand{\bracket}[1]{\left\langle\, #1\,\right\rangle}
\def\mat#1{\matt[#1]}
\def\matt[#1,#2,#3,#4]{\left(%
\begin{array}{cc} #1 & #2 \\ #3 & #4 \end{array} \right)}
\def\hq{\hat{q}}
\def\hp{\hat{p}}
\def\hx{\hat{x}}
\def\hk{\hat{k}}
\def\hw{\hat{w}}
\def\hl{\hat{l}}

\def\bea#1\ena{\begin{align}#1\end{align}}
\def\nn{\nonumber\\}
\def\cL{{\cal L}}
\def\TM{TM\oplus T^*M}
\newcommand{\CouB}[2]{\left\llbracket #1,#2 \right\rrbracket}
\newcommand{\pair}[2]{\left\langle\, #1, #2\,\right\rangle}

\null \hfill Preprint TU-968  \\[3em]
\begin{center}
{\LARGE \bf{Generalized geometry and nonlinear realization of 
generalized diffeomorphism on D-brane effective action}}
\end{center}

\begin{center}
{T. Asakawa${}^\sharp$\footnote{
e-mail: asakawa@maebashi-it.ac.jp}, H. Muraki${}^{\flat}$\footnote{
e-mail: hmuraki@tuhep.phys.tohoku.ac.jp}, S. Sasa${}^{\flat}$\footnote{
e-mail: sasa@tuhep.phys.tohoku.ac.jp}, and S. Watamura${}^{\flat}$\footnote{
e-mail: watamura@tuhep.phys.tohoku.ac.jp}}\\[3em] 
${}^\sharp$
Department of Integrated Design Engineering,\\
Faculty of Engineering,\\
Maebashi Institute of Technology\\
Maebashi, 371-0816, Japan \\[1em]

${}^\flat$
Particle Theory and Cosmology Group \\
Department of Physics \\
Graduate School of Science \\
Tohoku University \\
Aoba-ku, Sendai 980-8578, Japan \\ [5ex]

\abstract{The characterization of the DBI action of a $D_p$-brane using the generalized geometry is discussed.
It is shown that the DBI action is invariant under the diffeomorphism and $B$-transformation of the generalized tangent bundle of the target space.
The symmetry is realized non-linearly on the fluctuation of the $D$-brane.
}
\end{center}

\noindent{\sl\small Based on talk given by S.W. at the Workshop on Noncommutative Field Theory and Gravity, Corfu, Greece,
September 8-15, 2013.}

\vskip 2cm

\eject
\section{Introduction}

It is known that when a symmetry in the high energy theory is spontaneously broken to a smaller symmetry,
the low energy effective theory(LEET) is strongly controled by the symmetry breaking pattern.
If the broken symmetry is a continuous symmetry, LEET is described by the Nambu-Goldstone boson \cite{Nambu,Goldstone}
and the original symmetry in the high energy is realized nonlinearly which characterizes the LEET action.
This type of argument can also be applied to analyse the low energy D-brane action\cite{12051385}.
It is known that the D-brane effective action is given by DBI action.   
It was argued that the DBI action can be characterized also as a system with the spontaneous symmetry breaking
without reffering to the detail of the string theory \cite{Gliozzi,Casalbouni}.
There the Poincare symmetry of the target space plays the role of the high energy symmetry
and thus they are realized nonlinearly. 
In their formulation, the scalar boson which describe the 
transvers fluctuation of the brane are the Nambu-Goldstone modes,
and on the contrary the gauge boson on the brane is a covariant field 
of the nonlinear realization of the 
Poincar\'e group of the target space. 
The vector field is thus not considered as a Nambu-Goldstone mode
\cite{Gliozzi, Casalbouni}.

In this talk, I will show that the DBI action has an even bigger symmetry in the target space and 
there the vector field also appears as a Nambu-Goldstone mode. 
This becomes possible if we identify the D-brane as a Dirac structure in the generalized tangent bundle
of the target space. We also discuss other consequences on the effective theory of the
D-brane obtained from our formulation based on the generalized geometry.
This talk is based on the papers \cite{ASW, AMW}.


\section{Generalized geometry and D-brane}
\label{section2}

\subsection{Generalized geometry}
First let us introduce some concepts appearing in Hitchin's generalized geometry\cite{Hitchin, Gualtieri},
which we need to discuss our formulation.
In generalized geometry, instead of the tangent bundle $TM$ of $M$, 
we consider the generalized tangent bundle
$TM\oplus T^*M$. Its section is a generalized tangent vector which is denoted by
$u+\xi$ where $u\in\Gamma(TM)$ and $\xi\in \Gamma(T^*M)$. The Lie bracket is 
generalized to the Dorfman bracket which is not antisymmetric, 
and thus the Lie algebroid of the tangent vectors is 
generalized to a Courant algebroid.
As the symmetry of the Lie bracket is the diffeomorphism, the symmetry of the Dorfman bracket
is called the generalized diffeomorphism which is the diffeo.$\times$ B-transformation 
generated by a vector $u$ and a closed
$2$-form $\omega\in\Omega^2_{closed}$. 
The generator of the generalized diffeomorphism is a generalized Lie derivative ${\cal L}_{(u,\omega)}$.
A Dirac structure is a subbundle $L\subset TM\oplus T^*M$
of rank $D$ such that 
\begin{itemize}
\item Isotropic: for all $a,b\in \Gamma(L)$~, $\langle a,b\rangle=0$,
\item Involutive: $L$ is closed under the Dorfman bracket,
\end{itemize}
where $\langle a,b\rangle$ is the canonical inner product.
The Dirac structure defines a
Lie algebroid. 


\subsection{D-brane as a Dirac structure}

The reason why we identify the D-brane with the Dirac structure is the following.
A D-brane is a hyper surface accompanied by a line bundle.
Thus, it is described by the 
embedding map $\varphi$ from the $p+1$ dimensional world volume $\Sigma$ 
to the $D$-dimension target space $M$ (in the following we consider $M={\bf R}^D$ for simplicity),  
\[
\varphi ~:~ \Sigma\ni \sigma^a~ \rightarrow~ x^\mu(\sigma)\in M~,
\]
where $\sigma^a$ ($a=0,\cdots, p$) are brane coordinates and  $x^\mu$ ($\mu=0,1,\cdots,D-1$) are  coordinates of the 
target space.
We take the static gauge and denote the brane coordinates by $x^a=\sigma^a$ ($a=0,\cdots,p$)
and $x^i(\sigma)=0$.
The fluctuation of the $D$-brane is then given by scalar fields $\Phi^i(x)$($i=p+1\cdots D-1$)
corresponding to the transverse displacements and the gauge field
$A_a$ corresponding to the connection of the line bundle.

First, a Dirac structure defines a singular foliation on the target space $M$.
For example,  the Dirac strucutre given by
\begin{equation}
L_p=span\{\partial_0,\partial_1,\cdots,\partial_p,dx^{p+1},\cdots,dx^{D-1}\}~,
\end{equation}
defines a foliation, a leaf of which is a $p+1$ dimensional hyperplane.
We can identify a leaf as a D-brane then the translational invariance of the
foliation is broken, generating the Nambu-Goldstone modes.

The second reason is that the fluctuation can also be incorporated in the same manner.
When we take the Dirac structure $L_p$ to describe the Dp-brane in static gauge, 
the fluctuation of the Dp-brane, given by a vector field $A_a$ and scalar fields $\Phi^i$,
can be unified into a generalized connection ${\cal A}\in L_p^*$,
\begin{equation}
{\cal A}=A_a dx^a+\Phi^i\partial_i.
\end{equation}
where $L_p^*$ is the dual Dirac structure of $L_p$.
Then, the generalized field strength ${\cal F}$ of ${\cal A}$ is given by
\begin{equation}
{\mathcal F} = \frac{1}{2} F_{ab} dx^a \wedge dx^b + \partial_a \Phi^i dx^a \wedge \partial_i \quad \in \Gamma(\wedge^2 L^\ast),
\end{equation}
and it defines the deformed Dirac structure as
\begin{equation}
L_{\cal F}=e^{\cal F}L_p=span\{\partial_a + \partial_a\Phi^i \partial_i + F_{ab}dx^b,dx^i - \partial_a \Phi^i dx^a\},
\end{equation}
together with the condition that $d{\cal F}=0$.



\subsection{Symmetry of the Dirac structure}

Since the generalized diffeomorphism is given by diffeo.$\times$ $B$-field gauge transformation,
its generator is a generalized Lie derivative 
given by $\cal L_{\epsilon+\lambda}$. 
It transforms the deformed Dirac structure $L_{\cal F}$, but
the effect can be absorbed into the transformation of the fluctuation $\delta{\mathcal F}$ only.
This corresponds to keep the static gauge, and causes the non-linear transformation law for ${\mathcal F}$ 
as
\begin{subequations}
\begin{align}
\delta{\mathcal F}_{ab} &= (\partial_{[a}+ {\mathcal F}_{[a}^{\ \ j} \partial_j) (\lambda_{b]}-\epsilon^c {\mathcal F}_{cb]}-\lambda_k {\mathcal F}^k_{\ b]}) - (\epsilon^k - \epsilon^c {\mathcal F}_c^{\ k}) \partial_k {\mathcal F}_{ab}, \\
\delta{\mathcal F}_a^{\ j} &= (\partial_{a}+ {\mathcal F}_{a}^{\ \ i} \partial_i) (\epsilon^j - \epsilon^c {\mathcal F}_c^{\ j}) - (\epsilon^k - \epsilon^c {\mathcal F}_c^{\ k}) \partial_k {\mathcal F}_a^{\ j}.
\end{align}\label{eq:gene_Liederi_F:e47h}
\end{subequations}
When we evaluate them on the leaf of $x^i=\Phi^i(x^a)$, they correspond to the transformation law given by 
\begin{eqnarray}
\delta A_a &=& \lambda_a-\epsilon^c F_{ca}+\lambda_k\partial_a\Phi^k, \cr
\delta \Phi^i &=& \epsilon^i - \epsilon^c \partial_c \Phi^i.
\label{rule}
\end{eqnarray}
where we have imposed the static gauge condition on the fluctuation
and the coordinate $x^i$ in the parameter is replaced by the field $\Phi^i$.
As seen from this result, the transformations caused by $\epsilon^c$ and $\lambda_k$ are linear in fields.
Since they preserve the leaf, the unbroken symmetry is the worldvolume diffeomorphism and the $U(1)$-gauge transformation.
On the other hand, the inhomogeneous terms $\epsilon^i$ and $\lambda_a$ corresponds to the broken symmetry by specifying a particular leaf, and thus we can interpret that they are Nambu-Goldstone mode. 
Note that these transformation laws are extension of the non-linearly realized Poincar\'e symmetry in \cite{Gliozzi,Casalbouni}.
In particular, the term $\epsilon^i$ includes a 
translation along the transverse direction as well as a Lorentz rotation in a $a-i$ plane.

\subsection{Symmetry of the DBI action}

We have derived the non-linearly realized transformation rule of the fluctuation 
of the D-brane under the generalized diffeomorphism of the target space $M$. 
The DBI action is manifestly invariant under the worldvolume diffeomorphism on the D-brane and the $U(1)$-gauge transformation,
but it actually invariant under the full symmetry of the target space.
To see this invariance we first write the DBI action 
with the integration over the target space as
\begin{equation}
S_{\rm DBI} = \int_\mu  {\cal L}_{\rm DBI} \ \delta^{(D-p-1)}(x^i-\Phi^i(x^a)) \ dx^0 \wedge \cdots \wedge dx^{D-1},
\label{DBI2}
\end{equation}
where $\delta^{(D-p-1)}(x^i-\Phi^i(x^a))$ 
is a Dirac's delta function seen as a distribution along $x^i$-directions.
The Lagrangian is given by
\begin{equation}
{\cal L}_{\rm DBI} = \sqrt{\det (\varphi_\Phi^\ast (g+B) -F )_{ab}},
\label{DBIaction}
\end{equation}
where $g$  is the metric and $B$ the 2-form $B$ field,both defined in the target space evaluated on the leaf defined by the embedding map $\varphi_\Phi (\Sigma)$ of $L_{\cal F}$ at $x^i=\Phi^i (x^a)$.

By using the rule (\ref{rule}),
the transformation of the Lagrangian ${\cal L}_{\rm DBI}$ under the generalized diffeomorphism on the target space $M$ is given by
\begin{equation}
\delta {\cal L}_{\rm DBI} 
= -\epsilon^\mu \partial_\mu {\cal L}_{\rm DBI} -(\partial_c \epsilon^c +\partial_c \Phi^k \partial_k \epsilon^c){\cal L}_{\rm DBI}.
\label{Lie derivative of DBI}
\end{equation}
On the other hand, the delta function transforms as
\begin{align}
&\delta [\delta^{(D-p-1)}(x^i-\Phi^i)]\nonumber\\
=& -\epsilon^\mu  \partial_\mu  [\delta^{(D-p-1)}(x^i-\Phi^i)]
-(\partial_k\epsilon^k-\partial_k\epsilon^c\partial_c\Phi^k) \delta^{(D-p-1)}(x^i-\Phi^i).
\label{Lie derivative of delta} 
\end{align}
By combining them, we obtain 
\begin{equation}
\delta\left[{\cal L}_{\rm DBI} \,\delta^{(D-p-1)}(x^i-\Phi^i)\right]
= -\partial_\mu  \left[\epsilon^\mu {\cal L}_{\rm DBI} \, \delta^{(D-p-1)}(x^i-\Phi^i)\right] ~.
\end{equation}
The transformation of the integrand in the DBI action (\ref{DBI2}) is a total derivative and thus, 
 the DBI action is invariant under nonlinearly realized full target space diffeomorphisms and B-field gauge transformations.

\section{Discussion}

In the discussion, I want to talk about some other consequences
following the identification of the D-brane with a Dirac structure.
In the following, we consider the simplest Dirac structure, i.e. $L=TM$ which corresponds
to the spacetime filling D-brane. 
It is known that adding a $U(1)$ gauge flux $\omega$ as a fluctuation
describes a bound state of such a D-brane and lower dimensional branes.
As discussed in the previous sections, 
it defines a deformed Dirac structure $L_\omega$ provided $d\omega=0$, 
whose section has the form
\begin{equation}
V=v+\omega(v)~~,~~v\in \Gamma(TM),
\end{equation}
where $\omega(v)=i_v\omega$. 
In the generalized geometry, the same Dirac structure can also be
described as the deformation of the dual Dirac structure $T^*M$, as
\begin{equation}
V'=\xi+\theta(\xi)~~,~~\xi\in\Gamma(T^*M),
\end{equation}
where the $\theta$ is a Poisson 2-vector. 
We denote the corresponding Dirac structure by $L_\omega$.
The equivalence of the
two Dirac structures requires that there are always two
descriptions of the same vector, i.e. $V=V'$ and gives the relation
\begin{equation}
\theta=\omega^{-1}.
\end{equation}
We can also describe the generalized metric from these
two different Dirac structures.
The standard description is a description based on 
$TM$ and given by a tensor $E=g+B$ of $T^*M\otimes T^*M$. 
On the other hand, we can also describe the 
same metric structure from the Dirac structure $L_\theta$.
It is characterized by the tensor $t$ of $TM\otimes TM$.
The relation of the two tensors can be derived by the equivalence of the
two representations of the metric structure by
\begin{equation}
v+(g+B)(v)=\xi+\theta(\xi)+t(\xi),
\end{equation} 
and this gives the relation 
\begin{equation}
t+\theta=(g+B)^{-1}.
\end{equation}
Identifying $t$ with $G+\Phi\in \Gamma(TM\otimes TM)$, we get the relation appearing in the Seiberg-Witten map \cite{Jurco}.

The fluctuation on the D-brane bound state has also an alternative description.
In the standard description based on $L_\omega$, the fluctuation is a $2$-form $F$.
It can be considered as a deformation of the Dirac structure $L_{\omega +F}$, when $dF=0$.
Now from the Poisson side $L_\theta$, we can consider the fluctuation
by the deformation of the Poisson tensor $\theta$ by a 2-vector $\hat F$.
Then the generalized vector in the deformed Dirac structure $L_{\theta+\hat{F}}$ is given by 
\begin{equation}
V=\xi+(\theta+\hat F)(\xi).
\end{equation}
The condition that $L_{\theta+\hat{F}}$ is again a Dirac structure is now a Maurer-Cartan equation
\begin{equation}
[\theta,\hat F]_S+{1\over2}[\hat F,\hat F]_S=0,
\end{equation}
where $[\cdot,\cdot]_S$ is the Schouten bracket.
This is a Bianchi identity of a new type of the representation of $U(1)$ gauge theory.
In fact, we can also consider the $1$-vector potential and 
gauge transformation corresponding to this field strength.
Moreover, the explicit relation between two gauge potentials is obtained,
and the gauge-equivalence is shown in \cite{AMW}.
This type of gauge field may be interesting when we consider the 
non-geometric flux. See for example\cite{Blumenhagen}.

\section*{Acknowledgments}
Authors would like to thank 
the members of the particle theory and cosmology group, 
in particular U.~Carow-Watamura for helpful comments and discussions. 
T.~A., H.~M.and S.~S. are supported partially by the GCOE program ``Weaving Science Web beyond Particle-Matter Hierarchy'' 
at Tohoku University.
And H.~M. is also partially supported by Tohoku University
Institute for International Advanced Research and Education.

%

\end{document}